\documentclass{article}

\usepackage{arxiv}

\usepackage{graphicx}
\usepackage{subfig}
\usepackage{tabularx}
\usepackage{amsmath}
\usepackage[table, xcdraw]{xcolor} 
\usepackage{booktabs}
\usepackage[toc,page]{appendix}

\usepackage[utf8]{inputenc} % allow utf-8 input
\usepackage[T1]{fontenc}    % use 8-bit T1 fonts
\usepackage{hyperref}       % hyperlinks
\usepackage{url}            % simple URL typesetting
\usepackage{booktabs}       % professional-quality tables
\usepackage{amsfonts}       % blackboard math symbols
\usepackage{nicefrac}       % compact symbols for 1/2, etc.
\usepackage{microtype}      % microtypography
\usepackage{lipsum}
\usepackage{graphicx}

\title{Multi-scale structural complexity as a quantitative measure of visual complexity}

\author{
 Anna Kravchenko \\
  Radboud University, \\
 Nijmegen, The Netherlands\\
  \texttt{anna.kravchenko@ru.nl} \\
   \And
   Andrey A. Bagrov\\
  Radboud University, \\
 Nijmegen, The Netherlands\\
  \texttt{andrey.bagrov@ru.nl} \\
    \And
    Mikhail I. Katsnelson\\
  Radboud University, \\
 Nijmegen, The Netherlands\\
  \texttt{m.katsnelson@science.ru.nl} \\
    \And
    Veronica Dudarev\\
  University of British Columbia, \\
 Vancouver, Canada\\
  \texttt{vdudarev@mail.ubc.ca} \\
 }

\begin{document}
 \maketitle    
\begin{abstract}
While intuitive for humans, the concept of visual complexity is hard to define and quantify formally. We suggest adopting the multi-scale structural complexity (MSSC) measure, an approach that defines structural complexity of an object as the amount of dissimilarities between distinct scales in its hierarchical organization. In this work, we apply MSSC to the case of visual stimuli, using an open dataset of images with subjective complexity scores obtained from human participants (SAVOIAS). We demonstrate that MSSC correlates with subjective complexity on par with other computational complexity measures, while being more intuitive by definition, consistent across categories of images, and easier to compute. We discuss objective and subjective elements inherently present in human perception of complexity and the domains where the two are more likely to diverge. We show how the multi-scale nature of MSSC allows further investigation of complexity as it is perceived by humans.
\end{abstract}

\maketitle

\section{Introduction}

In studying information processing in human perception, attention, and thinking, we are often faced with the issue of characterising the information to be processed in objective terms. One of the earliest steps in perception research - what we now know as psychophysics - aimed at discovering the laws that connect perceived properties of stimuli to their physical properties, such as brightness of a dot, loudness or pitch of a sound. The relationships between subjectively perceived and objective complexity could be studied as well, and could inform studies on downstream effects of complexity, e.g. on aesthetics, attention, and motivation \cite{Berlyne1960,deWinter2023,Kondyli2023}.

For example, visual complexity is an important factor in determining aesthetic preference for artistic works \cite{Forsythe2011,Jacobsen2002}. Early works suggested a linear relation between visual complexity and aesthetic beauty  \cite{Eysenck1941, Reinecke2013}, where perceived beauty is proportional to aesthetic order but decreases with increased complexity \cite{Birkhoff1933}. However, later studies revealed that the relationship is more of an optimum type: most appealing images tend to have intermediate levels of visual complexity \cite{Berlyne1971}.

Curiosity is another field where complexity of information is an important factor, along with novelty and learning difficulty  \cite{Berlyne1960,Kidd2015}. While curiosity was shown to improve learning speed and outcomes in AI systems \cite{Twomey2017,Gottlieb2018,Oudeyer2018}, AI models implementing curiosity measures relying on complexity and novelty seem to be disrupted by noise \cite{Burda2018,Oudeyer2018}, and more recent research sees the preference for intermediate complexity as coincidental with optimising learning progress \cite{Poli2024}. 

Characterizing complexity objectively, however, has been a persistent issue, not only in psychology. First attempts to quantify complexity arose in coding theory and data compression, focusing on statistical properties of data and the amount of information needed to reliably transmit it in a message. Shannon, \cite{Shannon}, quantified information through entropy, the amount of uncertainty present in a message, while Kolmogorov defined complexity as the length of the description needed to reproduce the stimulus \cite{Kolmogorov, Sun2022}.

\begin{figure}[hbt!]
  \centering
  \includegraphics[width=16cm]{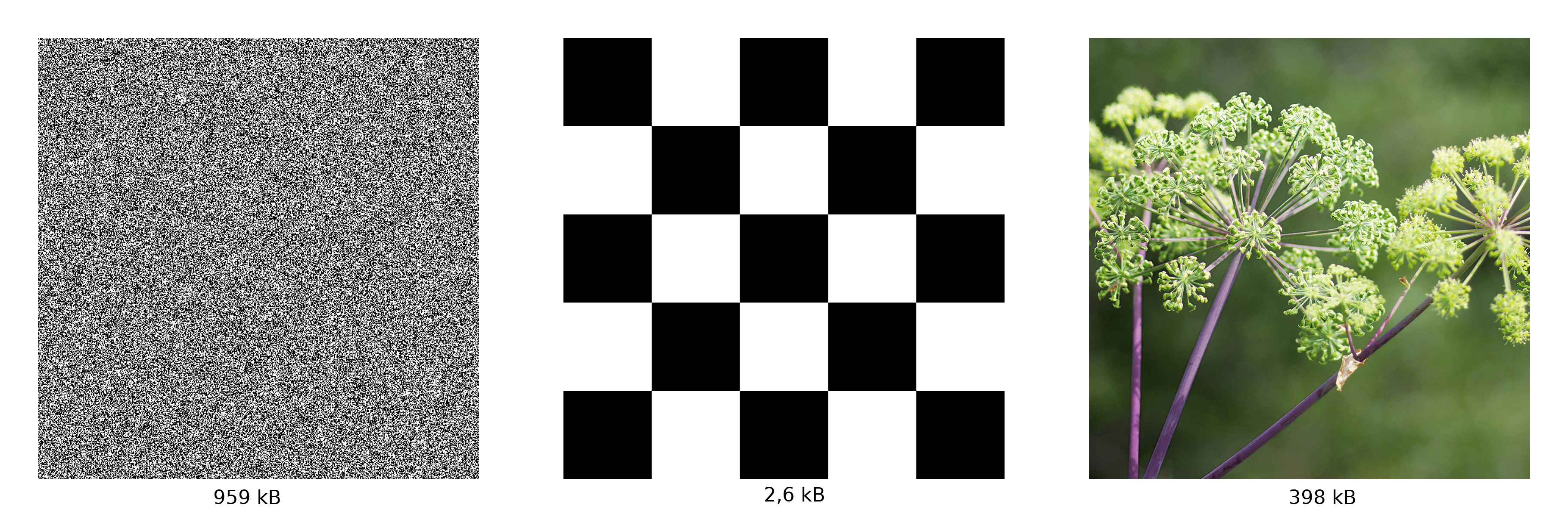}

  \caption{Descriptive measures of complexity tend to prioritize pure randomness over structure, ranking noise as more complex than meaningful images.}
\label{image_examples}
\end{figure}

These measures are usually referred to as measures of informational complexity and they focus on the amount of randomness present in data rather than structural nontriviality. There are multiple ways to implement these concepts. For example, file size \cite{Machado2015} can be viewed as an upper bound on Kolmogorov's complexity of its content. To an extent, these measures do correlate with subjective ratings of complexity provided by human observers \cite{deWinter2023,Machado2015,Saraee2018}, however, they also assign high complexity to random stimuli (Fig.~\ref{image_examples}, left), which intuitively are not perceived as complex. Meanwhile, things that we humans perceive as truly complex tend to have a balance between order and randomness (Fig.~\ref{image_examples}, right), i.e. they are structurally nontrivial. 

In psychological literature, several qualitative analyses of complexity describe it as a function of three factors: the number of elements (which is still consistent with Kolmogorov and Shannon), the dissimilarity between the elements, and their organisation within the stimulus \cite{Berlyne1960, Donderi2006, Oliva2004}. Dissimilarity and organisation of items within a stimulus proved to be hard to quantify computationally. Edge density and visual clutter \cite{Oliva2004_edge, Rozenholz2007} are some examples of the attempts to do so. However, defining dissimilarity and especially organization in a way that is generalizable across different types of stimuli is challenging \cite{Donderi2006}. One of the most successful attempts to quantify this aspect of complexity is a measure of the "number of regions" \cite{Comaniciu2002}. This computational measure correlates quite well with subjective complexity \cite{Saraee2018}, especially in combination with other measures \cite{Nuthmann2015}. Its downside is that it requires some user input in setting the parameters, without clear guidelines on how to do that. 

So far we have reviewed the attempts to quantify complexity as a single variable. \cite{Oliva2004} suggested that complexity is not a uni-dimensional metric. Yet to the best of our knowledge no computational model of dimensional measure of complexity has been proposed. In some way, multi-scale structural complexity can be considered as a first step in this direction, as we show below.

Another persistent issue in quantifying complexity is the relationships between objective and subjective complexity. While developers of computational measures of complexity aim at perfect correlation with subjectively rated complexity, one could theorise that perfect agreement is - and should not be - possible. \cite{Madan2018} showed that human ratings of complexity are inherently influenced by affective value of the information. That is, what we see as more emotional is rated higher on complexity. Another factor in human perception of complexity is recognisability - or familiarity - of the pattern. The best example of this is characters in a language one can versus cannot read. A mathematical formula, while being a fairly simple visual pattern might convey incredibly detailed information. Recognisability of the pattern, being a property of the perceiver, not the perceived, is impossible to quantify objectively. Together, these examples suggest that perceived complexity does not reflect just the number, dissimilarity, and organisation of the elements, but includes additional factors. These additional factors could be the reason why most of the existing measures of objective complexity show different magnitude of correlation with subjective complexity for photographs as compared to paintings, the latter presumably having more prominent emotional and cultural aspects \cite{Saraee2018}. They go as far as to conclude that different computational measures of complexity are more suitable for different domains (stimulus types).

The notion of complexity also happens to be important in physics where it is used to describe emergent phenomena in systems of multiple components \cite{Bagrov2020}. In this work, we use a measure called multi-scale structural complexity (MSSC), and apply it to the case of visual stimuli. We compare it to other computational measures of complexity and subjective complexity, and demonstrate how its multi-scale nature can help investigate human perception of image complexity.

\section{Multi-scale structural complexity}

One of the hallmarks of the majority of complex systems observed in the world -- from biological structures to pieces of art -- is the co-existence of a number of well-defined characteristic scales. In other words, most complex systems have hierarchical organisation. For example, any living organism is structured in a multi-level way, with levels of organs, tissues, cells, sub-cell organelles, and complex molecules reaction networks being fundamentally different from each other. It was suggested that competing interactions between these levels is what gives raise physical and biological complexity \cite{Wolf2018}. Similarly, an image (a painting or photograph) has an overall gist, separable objects, and fine detail content. 

The idea of complexity as self-dissimilarity at different spatial or temporal levels has been embraced in \cite{Wolpert1997} and \cite{Wolpert2007}. Following this line of thinking, the concept of multi-scale structural complexity (MSSC) has been introduced \cite{Bagrov2020}. MSSC originated in physics and was first used to describe phase transitions in classical and quantum systems composed of many components. At the same time, since it was inspired by the intuitive human perception of complexity, applying it in the realm of visual perception is a natural endeavour, which we pursue in this paper. 

MSSC quantifies the amount of distinct scales present in a visual pattern using the idea of coarse graining, or Renormalisation Group (RG), borrowed from physics. The formal mathematical definition of MSSC is provided in the Appendix. Here we briefly explain how it works on a more conceptual level. Step by step, information is erased from the pattern -- first from the most detailed microscopic scale, and then from larger and larger scales, as shown in Fig. 2. This generates a stack of patterns ${\cal P}_i$ derived from the original image (denoted as ${\cal P}_0$). Now, assume that we compare two subsequent patterns in this stack, ${\cal P}_k$ and ${\cal P}_{k+1}$. If the difference between them is substantial, it implies that considerable amount of information has been lost at the coarse graining step $k+1$. This, in accordance with the idea of multi-scale dissimilarity, would produce a large value of {\emph partial complexity} ${\cal C}_k$ to scale $k$. On the other hand, if ${\cal P}_k$ and ${\cal P}_{k+1}$ are nearly identical, scale $k$ does not bear any unique features, and complexity value associated with it is low. Cumulative sum of partial complexities over relevant scales ${\cal C} = \sum\limits_k {\cal C}_k$ is called multi-scale structural complexity.

\begin{figure}[hbt!]
  \centering
  \includegraphics[width=16cm]{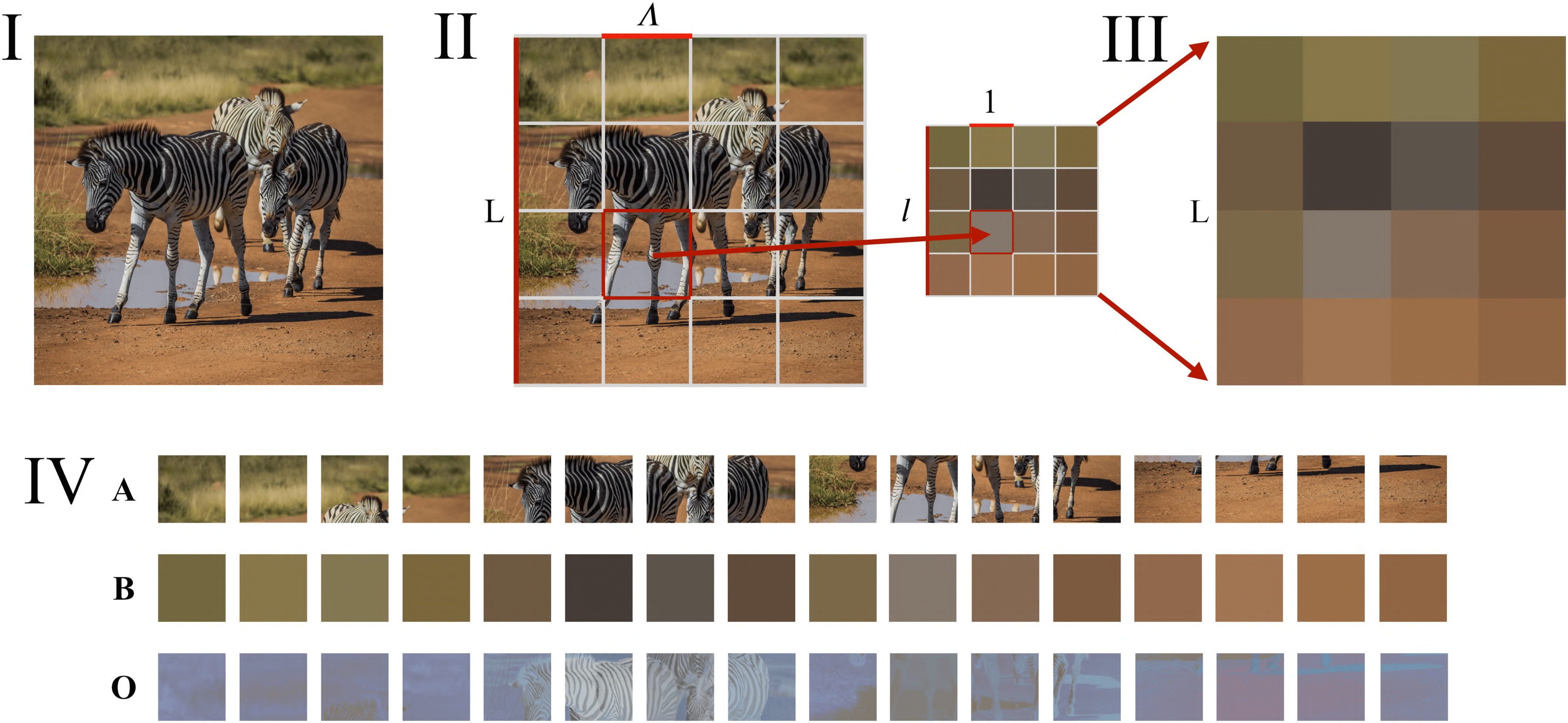}

  \caption{from (Bagrov et al., 2020): schematic representation of the idea behind the coarse graining method. The difference between versions of the image at each step of coarse graining determines partial complexity contributed by the spatial scale removed at that step.}
\label{mssc_old}
\end{figure}

MSSC should be regarded as an umbrella concept rather than a singular definition, as it can be implemented in multiple ways. First of all, the coarse graining procedure can be performed in a variety of ways. At this stage of MSSC development, the choice should be guided by theoretical considerations of which approach best suits the problem. For example, in the original paper, it has been shown that even something as simple as averaging over segments of a picture as shown in Fig.\ref{mssc_old} provides sufficient results when addressing the problem of phase transitions in complex physical systems. In the current work, focusing on visual complexity, we chose an approach that is biologically plausible for human visual perception (see details in the next section). 

The second aspect is that not all scales formally present in the pattern are relevant and have to be taken into account. In \cite{Bagrov2020}, it was shown that better results in identifying phase and structural transitions can be achieved if one neglects the smallest scales, where the very notion of structure and correlation length is not established yet, and the largest scales that exceed the maximal characteristic length of pattern features. In this paper, we will study how partial complexities of different scales of a visual stimulus correlate with the human ranking, and show that the best practice is indeed to account for a particular range of scales when computing $\cal C$.

\section{Implementation}

\begin{figure}[hbt!]
  \centering
  \includegraphics[width=16cm]{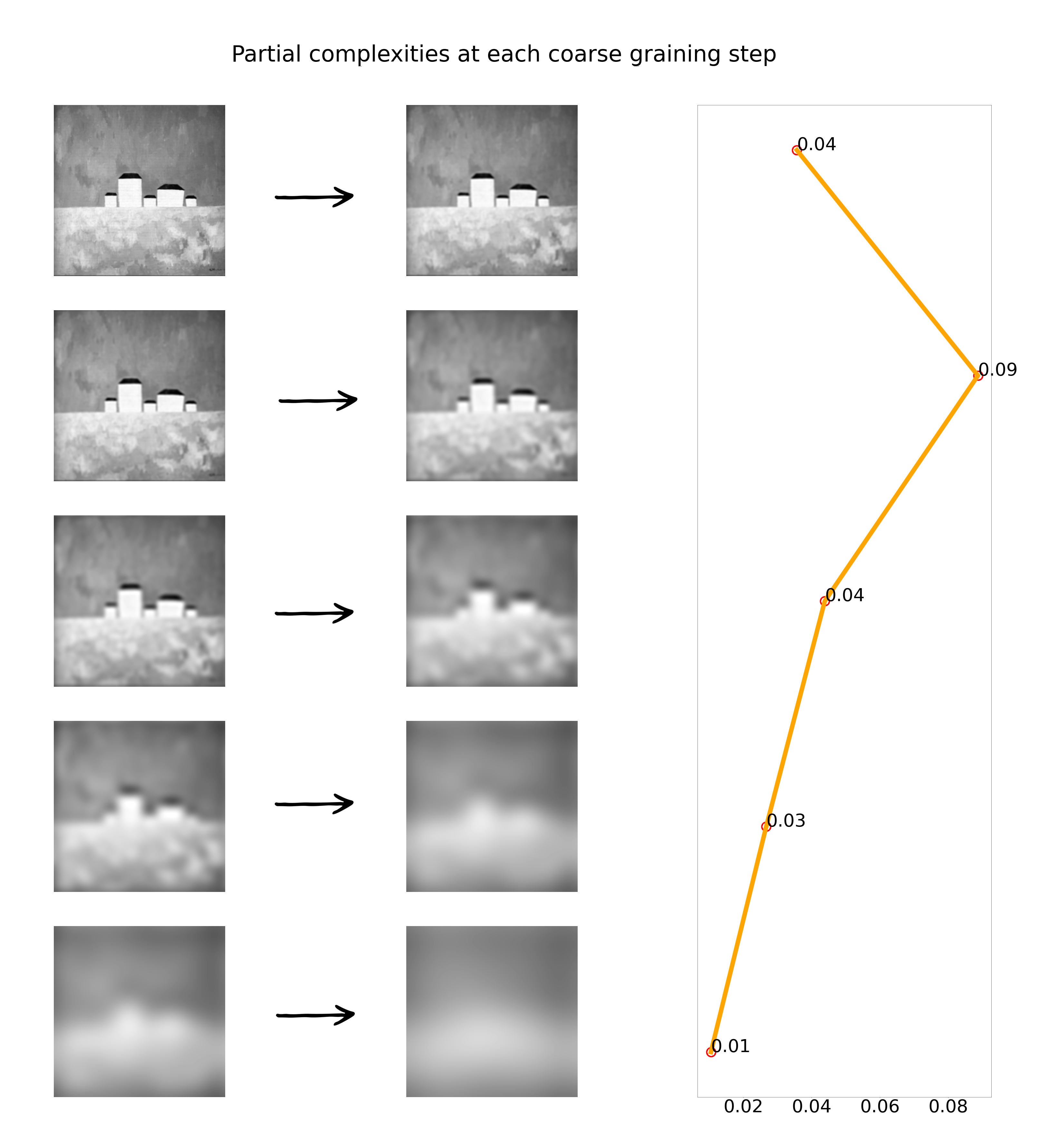}

  \caption{Coarse graining performed with Fourier Transform, the graph on the right shows partial complexity for each step. }
\label{coarse_graining_example}
\end{figure}

It is possible to implement coarse graining in a variety of ways, which allows us to  aim for the more biologically plausible approach. Evidence suggests processing on early layers of the visual cortex can be approximated by Fourier transform \cite{Campbell1968,Ochs1979,Kulikowski1981,Olshausen2003, Stevens2004,Kesserwani2020}. Fourier transform analyses a signal into frequencies that compile it. An image would be Fourier-transformed into a sum of spatial frequencies, where high-frequency components correspond to fine details, textures and edges, and low-frequency components correspond to larger shapes and smooth, gradual variations in intensity. Fourier transform is often used to characterize and manipulate visual stimuli in cognitive science \cite{Marr2010}. Here we use it to perform coarse graining. 

Namely, we decompose the image into spatial frequencies by applying discrete Fourier Transform (Eq.~\ref{DFT}). Then, step by step, we remove from the sum of the spatial frequencies the highest band (low pass filter) and reconstruct the image from what remains (Eq~\ref{inverse}): 
\begin{equation}
F_{k_x,k_y}=\sum_{n_x,n_y=0}^{N-1} f_{n_x,n_y} \cdot e^{-i 2 \pi \frac{k_xn_x +k_yn_y}{N}},
\label{DFT}
\end{equation}
\begin{equation}
\tilde{f}_{n_x,n_y}=\frac{1}{N^2} \sum_{k_x,k_y=0}^{N-K-1} F_{k_x,k_y} \cdot e^{i 2 \pi \frac{k_xn_x +k_yn_y}{N}},
\label{inverse}
\end{equation}
where $f_{n_x,n_y}$ is the intensity of pixels in the original grey-scale two-dimensional image (with $n_x,n_y$ pixel coordinates), and $\tilde{f}_{n_x,n_y}$ is its coarse-grained version obtained by removing $K$ highest frequencies. An example of the process can be seen in Fig.~\ref{coarse_graining_example}.

%In addition to Fourier transform, we tested several other approaches to coarse graining, including Gaussian blur and wavelet transform. Comparison of those falls beyond the scope of this paper, however, examples and reasoning behind choosing FT can be found in Supplementary material. The code for all attempted methods can be found \href{https://github.com/ankravchenko/image_complexity_analysis}{on Github}.

\section{The present study}

We used a published set of images with computational and subjective complexity measures - SAVOIAS \cite{Saraee2018} to estimate the correlation between their human ranked complexities and MSSC values. We selected this dataset because it is an open source one that provides access to subjective complexity value for each image. In addition, authors \cite{Saraee2018} supplement subjective ratings with computational measures of complexity, which allows us to compare MSSC to those as well.

\begin{figure}[hbt!]
  \centering
  \includegraphics[width=16cm]{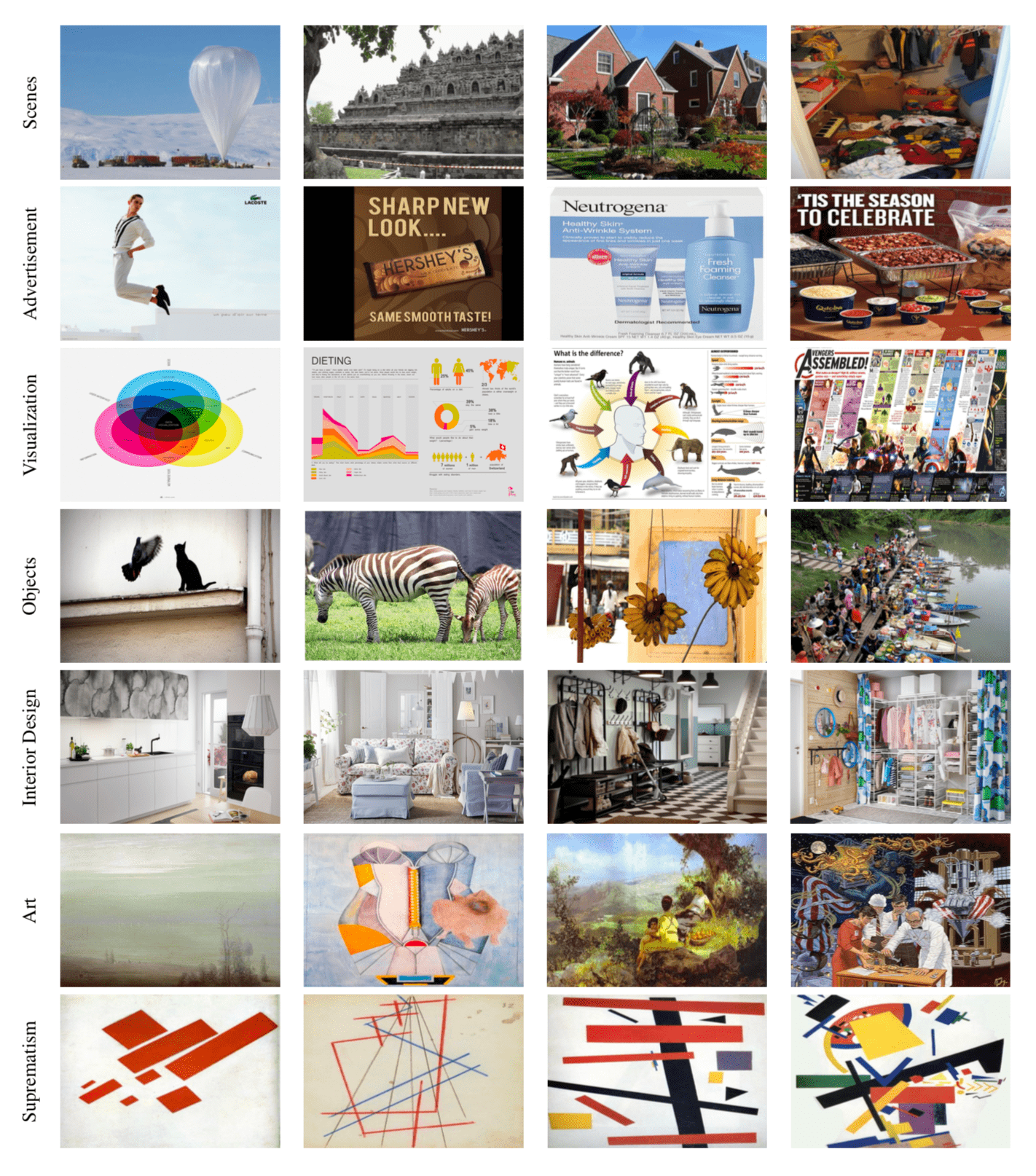}

  \caption{Categories in the SAVOIAS dataset (Saraee et al.,2018)}
\label{savoias}
\end{figure}

\subsection{Dataset}

SAVOIAS \cite{Saraee2018} is a set of 1420 images grouped into seven categories: Scenes (photographs of natural scenes), Advertisements, Visualization and Infographics, Objects (photographs of objects in natural context), Interior design (photographs of interior design displays from Ikea cataglogue), Art, and Suprematism (the distinction between Art and Suprematism was introduced by \cite{Saraee2018}, we keep this distinction for consistency). 

Subjective complexity for these images was obtained from 1,687 participants in the following way: Participants were presented with pairs of images and were asked to select the more complex one. Each participant was presented with a subset of images, all coming from the same category. The resulting relative scores were then converted to a continuous measure of visual complexity using Bradley-Terry method and matrix completion. For full details about the image selection, as well as subjective and objective measures of complexity, see \cite{Saraee2018}.

\subsection{Methods}

MSSC was computed in the way described above (see Implementation, the code is available \href{https://github.com/ankravchenko/image_complexity_analysis}{on Github}) for each image. We processed each colour channel separately, calculating its intensity before estimating complexity of this channel and then summed the resulting complexities multiplied by intensity for all channels. 

As we explained above, partial complexities at each step of coarse graining may have varying impact on overall complexity. They are also not equally meaningful. Elements present only at the smallest scales are likely image artifacts or details insignificant to the human eye, while later steps of coarse graining, at which objects start to disappear, happen on the scales comparable to the scale of image itself. In both cases each step of coarse graining adds significant change regardless of the content of the image.

To account for that in our analysis we only used the middle scales for measuring the total complexity of images and calculated the correlation between partial complexity at each scale and human ranking to ensure the correct cutoff points. 

We then removed images that were more than 2 standard deviations away from the category average on MSSC (such outliers were found only in the Art category, and we discuss them separately), and computed Pearson correlation between MSSC and perceived complexity provided by \cite{Saraee2018}.

\section{Results}
Figures 5 and 6 show scatterplots for the correlations between MSSC and subjective complexity for each category, which we further divided into two clusters: Natural scenes refer to images obtained by means of photography, while ``Man-made images'' refer to images painted or digitally produced by humans. Although this division is purely heuristic, the figures show that man-made images produce noisier correlations between MSSC and subjective complexity. This is especially evident for the "art" category, where MSSC assigns a high value to what humans perceive as not very complex. We will return to this point in the discussion. 

\begin{figure}[hbt!]
\centering
\includegraphics[width=18cm]{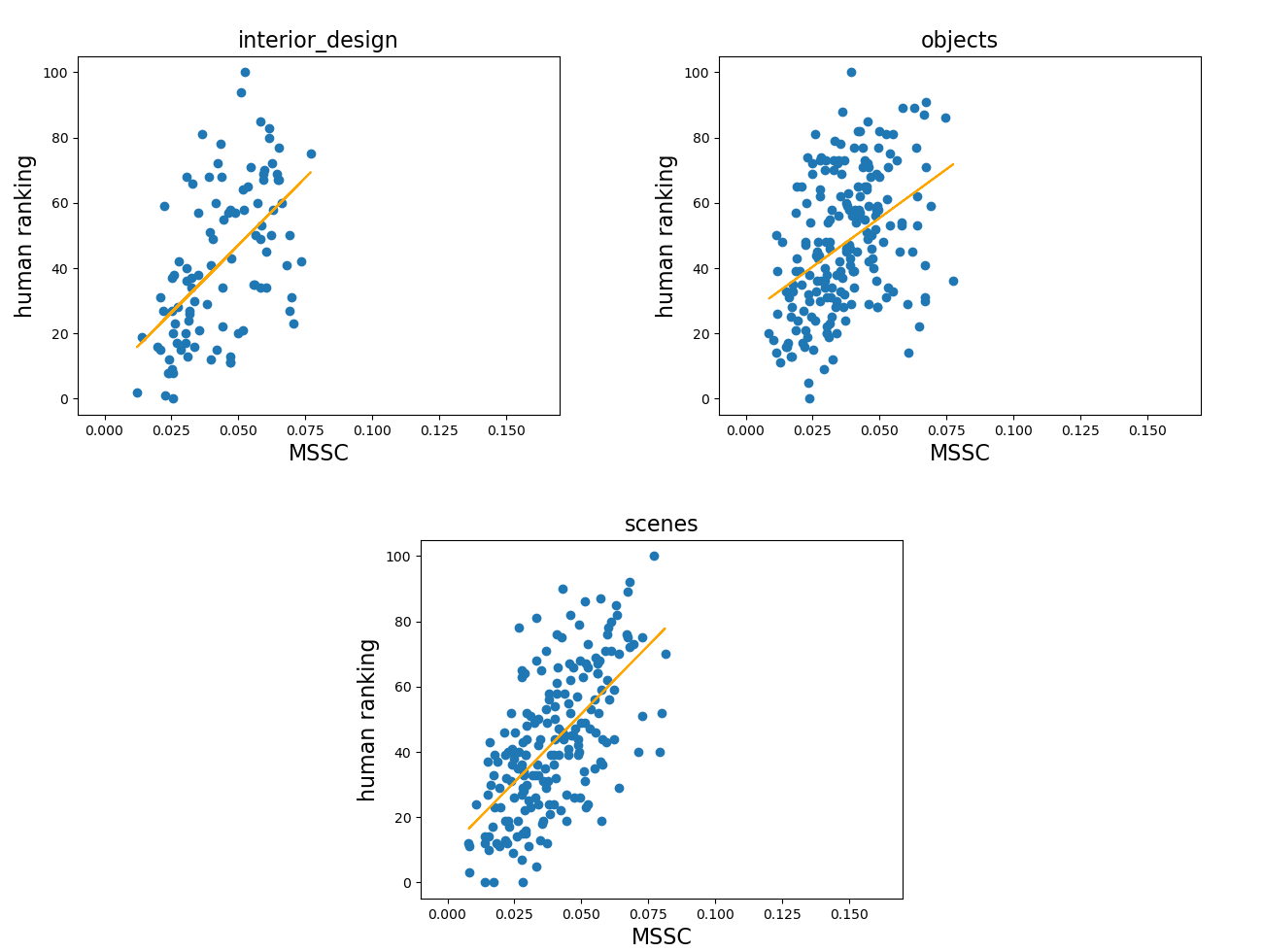}

\caption{Natural scenes}
\label{natural}
\end{figure}

\begin{figure}[hbt!]
\centering
\includegraphics[width=18cm]{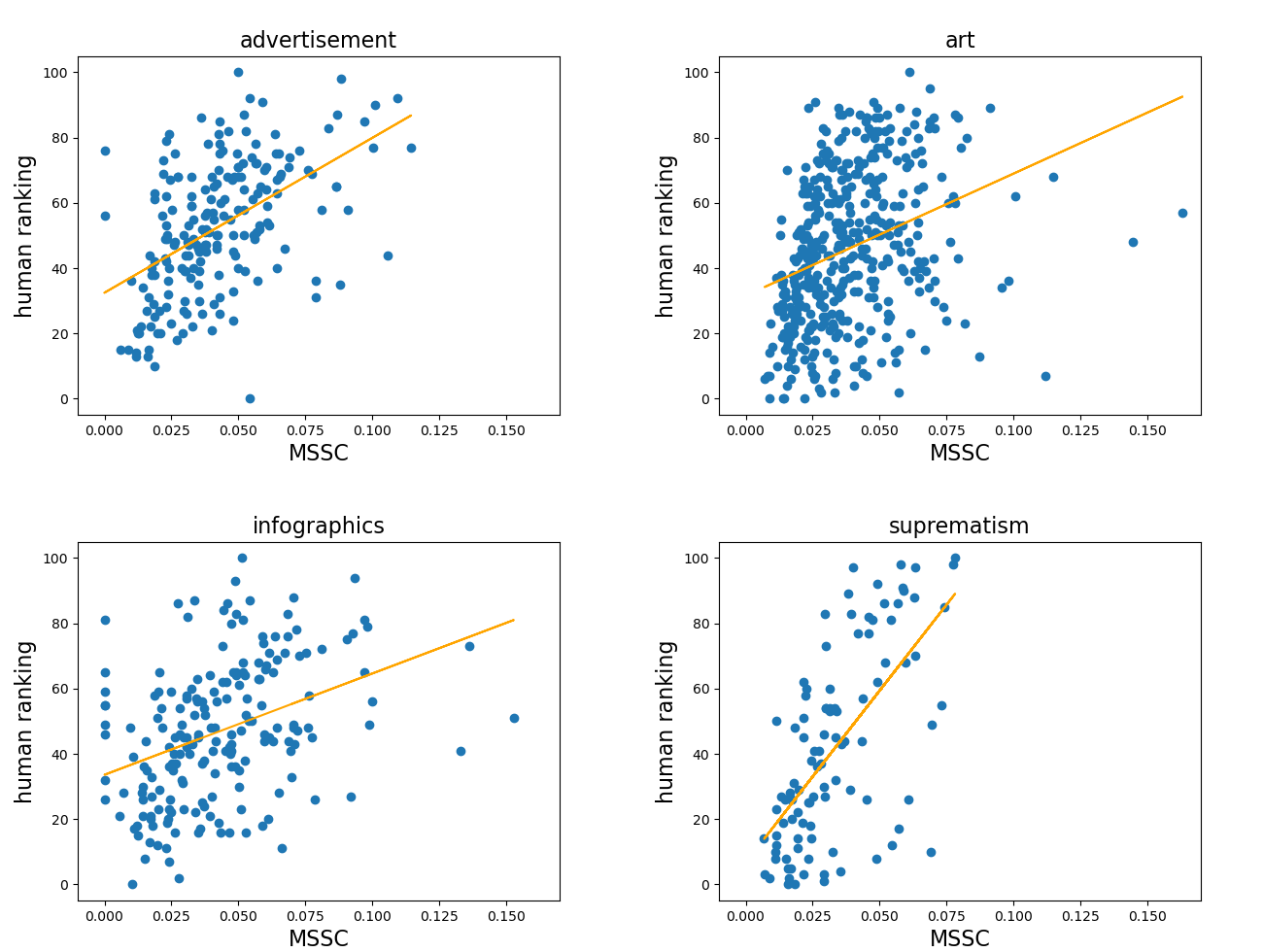}
\caption{Man-made images}
\label{artificial}
\end{figure}

\newpage
Table 1 shows Pearson correlation between MSSC and subjective complexity (first column). It also shows correlations from Saraee et al., 2018, between subjective complexity and other computational measures of complexity.

\begin{table}[hbt!]
\caption{Pearson correlations between subjective complexity and computational measures of complexity: MSSC as computed in the present work, edge density and other measures taken from Saraee et al., 2018.} %The highest correlation (numerically) is highlighted in bold font.}
\label{comparison}
\begin{tabular}{|l|c|c|c|c|c|c|}
\hline
 & edge density & compression ratio & number of regions & feature congestion & subband entropy & MSSC\\ \hline
\rowcolor[HTML]{FFFFFF} 
\cellcolor[HTML]{D9EAD3}scenes & 0.16 & 0.3 & 0.57 & 0.42 & 0.16 & \cellcolor[HTML]{D9EAD3}0.62  \\ \hline
\rowcolor[HTML]{FFFFFF} 
\cellcolor[HTML]{D9EAD3}objects & 0.28 & 0.16 & 0.29 & 0.3 & 0.1 & \cellcolor[HTML]{D9EAD3}0.46 \\ \hline
\rowcolor[HTML]{FFFFFF} 
\cellcolor[HTML]{FFF2CC}suprematism & 0.18 & 0.6 & \cellcolor[HTML]{D9EAD3}0.84 & 0.48 & 0.39 & \cellcolor[HTML]{FFF2CC}0.76  \\ \hline
\rowcolor[HTML]{FFFFFF} 
\cellcolor[HTML]{FFF2CC}interior design & 0.61 & 0.68 & \cellcolor[HTML]{D9EAD3}0.67 & 0.58 & 0.31 & \cellcolor[HTML]{FFF2CC}0.60 \\ \hline
\rowcolor[HTML]{FFFFFF} 
\cellcolor[HTML]{FFF2CC}advertisements & 0.54 & \cellcolor[HTML]{D9EAD3}0.56 & 0.41 & 0.56 & 0.54 & \cellcolor[HTML]{FFF2CC}0.52 \\ \hline
art & \cellcolor[HTML]{FFFFFF}0.48 & \cellcolor[HTML]{FFFFFF}0.51 & \cellcolor[HTML]{D9EAD3}0.65 & \cellcolor[HTML]{FFFFFF}0.22 & \cellcolor[HTML]{FFFFFF}0.33 & 0.36  \\ \hline
infographics & \cellcolor[HTML]{FFFFFF}{\color[HTML]{274E13} 0.57} & \cellcolor[HTML]{FFFFFF}0.55 & \cellcolor[HTML]{FFFFFF}0.38 & \cellcolor[HTML]{FFFFFF}0.52 & \cellcolor[HTML]{FFFFFF}0.61 & 0.38 \\ \hline
\end{tabular}
\end{table}

\begin{figure}[hbt!]
  \centering
  \includegraphics[width=14cm]{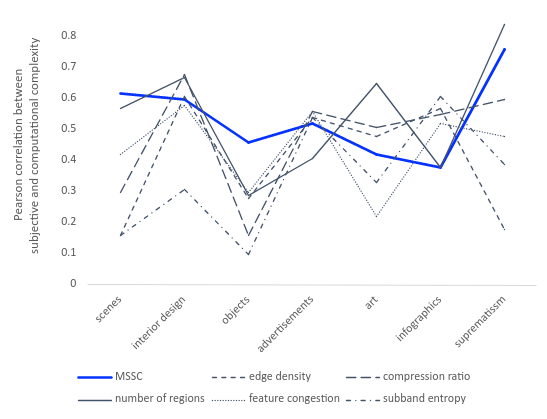}

  \caption{Pearson correlations between subjective complexity and computational measures of complexity, visualised. We can see that MSSC (blue line) behaves more consistently across different domains.}
\label{comparison_vis}
\end{figure}

\newpage

MSSC outperformed all the other computational measures for the Scenes and Objects categories (Table \ref{comparison}, Fig.~\ref{natural}). For the Scenes, Pearson's correlation between MSSC and subjective complexity was 0.62, which was close, but slightly higher than the maximal 0.57 achieved by the other computational measures (number of regions). Objects were the most challenging category for all computational measures, with the maximal correlation being just 0.3, while MSSC reached 0.46. 

For Suprematism, Interior Design and Advertisement, MSSC performed on par with the other existing measures.

The poorest performance, relative to the existing measures, was observed for Art and Infographics. Notably, those are the most information-driven, symbolic images prone to more variable subjective interpretations. We return to this point in the discussion. It is noteworthy that, descriptively speaking, MSSC agreement with subjective complexity varied the least between the different categories.%, as shown in Figure \textcolor{red}{(what I sent in the email) - shall we include it?}.

Taken together, these results suggest that MSSC as a measure of visual complexity performs on par or sometimes better than the other computational measures considered here. Yet there is considerable mismatch between MSSC and subjective complexity, and the amount of this mismatch varies across categories of images. To investigate these mismatches, we will look at where complexity arises in the organisational hierarchy of the image.

%Man-made images presented a greater challenge. The poorest performance, relative to existing measures, was observed for categories with more information-driven, symbolic imagery or pictures prone to more subjective interpretations, namely infographics and art (Pearson correlation of 0.38 and 0.42, compared to the best performance of 0.61 and 0.65). Conversely, MSSC demonstrated competitive performance among other man-made categories such as interior design, advertisement, and suprematism, approaching the best existing measures. Images in these categories might require culturally specific knowledge to interpret them fully, however, they still rely more on information conveyed directly through perceptual properties rather than through symbolic relationships.

%These results demonstrate MSSC's potential as a versatile and valuable approach across most image types, sparing the need to use a different measure for each as \cite{Saraee2018} suggested. They also imply that we are trying to measure at least two separate things as a singular complexity value, which may not be the optimal approach. 

\subsection{Where in the hierarchy does complexity arise}

\begin{figure}[hbt!]
  \centering
  \includegraphics[width=14cm]{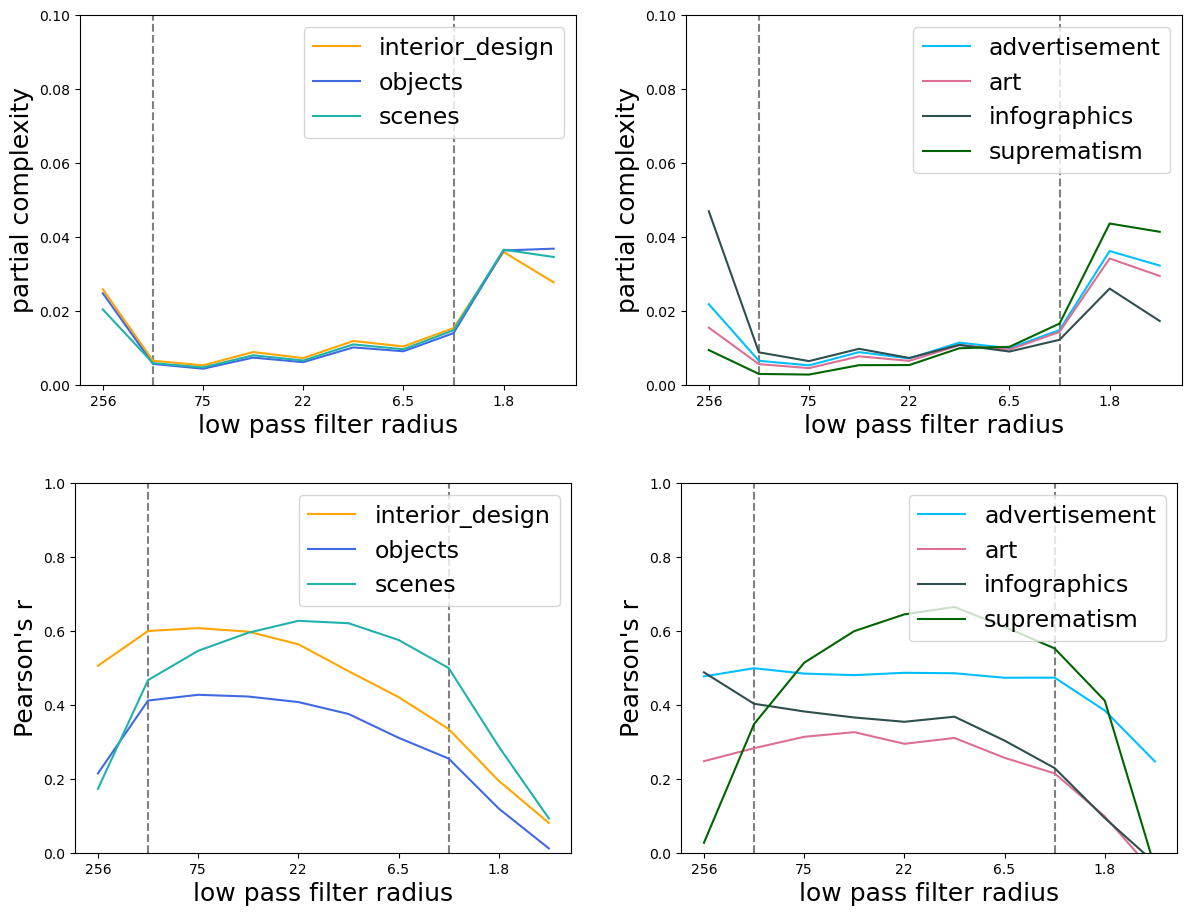}

  \caption{Partial complexity at each step of coarse graining (top row) and its correlation with subjective complexity (bottom row) for the different categories (shown in color). Low pass filter radii for coarse graining steps were spaced evenly on a log scale.}
\label{scale_impact}
\end{figure}

A unique feature of MSSC is that it allows and even invites investigation of where complexity arises in the hierarchy of the scales. This can be quantified as the magnitude of partial complexity for each step of the MSSC procedure (Equation S5, Fig. \ref{scale_impact}, top raw). Partial complexity can then be correlated with subjective complexity (Fig. \ref{scale_impact}, bottom raw). 

As can be seen from Fig. \ref{scale_impact}, the middle range of spatial frequencies had the lowest partial complexity, while the very high and the very low frequencies had higher partial complexity, and thus had the strongest impact on the overall MSSC. Yet the correlation with subjective complexity showed a reversed pattern: it was lowest at the first and the last steps of MSSC. That is, spatial scales that matter the most for subjective complexity turned out not the same that had the most impact on the overall MSSC (Fig. \ref{scale_impact}). 

Fig. \ref{scale_impact} also shows different patterns for the different categories of images, especially regarding the impact and correlation with partial complexity at the lowest levels (fine-grained details).  %\textcolor{red}{Anything about truncation of scales?}

\subsection{Discrepancy with human ranking}

 A common problem for many computational measures of complexity is that sometimes they produce unreasonably high or low values. We therefore investigated whether there are common features of the images that correlated with human ranking the least.
 
\begin{figure}
    \centering
    \includegraphics[width=14cm]{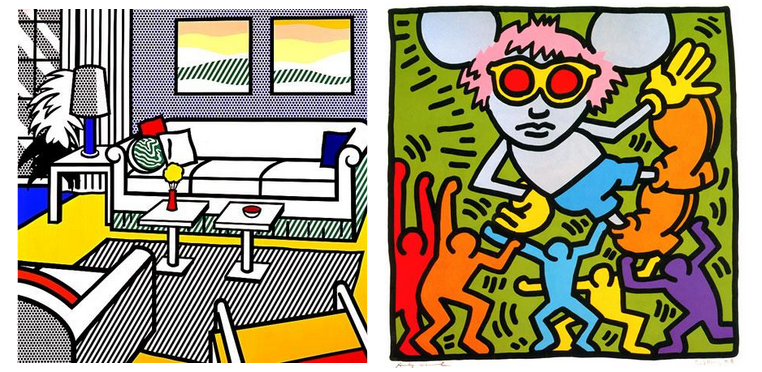}
    \caption{Some of the images from the Art category diverging from prediction by a threshold greater than two standard deviations. 
}
    \label{outliers}
\end{figure}

 First, we looked at the images that correlated the least with human ranking. Paintings diverging from prediction by a threshold greater than two standard deviations, were considered by MSSC more complex than eastimated by human participants and exhibited the same distinctive visual characteristic: broad lines featuring contrasting colors. Some examples are shown in (\ref{outliers}). This artistic style can be visually straining, however, it's relatively simplistic in terms of information conveyed through it.

\begin{figure}[hbt!]
  \centering
  \includegraphics[width=14cm]{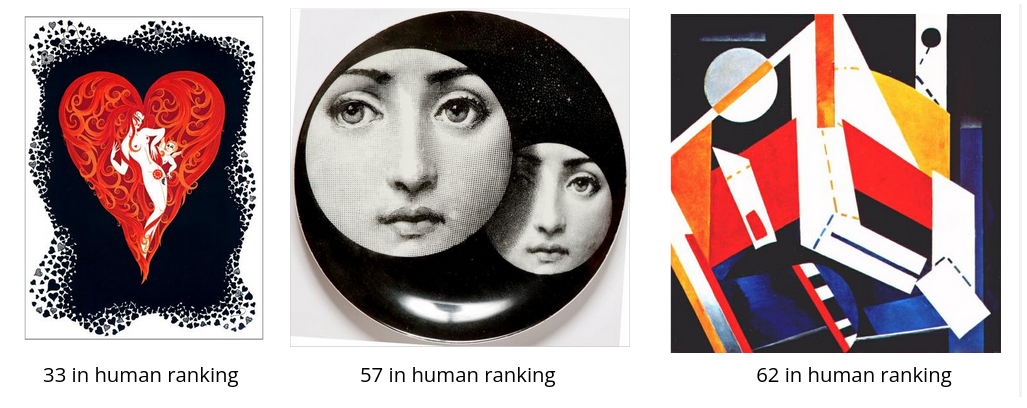}
  \caption{Images with the same MSSC ranking that were ranked vastly differently by human participants. We attribute the difference to subjective estimates of artistic value and "message" conveyed by the picture.}
\label{subjectivity}
\end{figure}

To further explore the differences between subjective complexity and MSSC we looked at images with identical MSSC values and huge discrepancy between human participants (\ref{subjectivity}). This discrepancy could be explained by subjective estimates of artistic value of the picture and information conveyed. Similar tendencies can be found in the infographic set.

%Full analysis with the list of outliers and examples of images with disparate rankings can be found in the Supplementary material.

\section{Discussion}

%\subsection{Performance and the definition of complexity}

We have found significant correlations between MSSC and subjective complexity. MSSC performs on par with other computational methods in most cases, surpassing them on natural images and providing more consistent results across categories. 

Having said that, it is evident that MSSC was not equally correlated with subjective complexity across categories. Moreover, the impact of the partial complexity at the lowest levels was numerically higher for natural than man-made scenes, yet its correlation with subjective complexity was lower. This pattern suggests that the more symbolic, culturally defined were the elements present in the picture, the higher was the correlation between smaller spatial scales and subjective complexity. It could be that human viewers pay more attention to details when they assume intentionality behind these details, or expect them to be informative. This, in turn, possibly drives a different way of expressing the details in paintings (art) and digital creations (advertisement, infographics). A striking example of this is shown in Fig. 8, which presents two extreme cases from the "art" category where MSSC was much higher than subjective complexity. In this case, MSSC was evidently overwhelmed by the small details, yet human participants largely ignored them in their complexity judgements. An interesting outlier among man-made images is suprematism art, where participants' distribution of attention to spatial frequencies mimics that for natural images. Should it be taken as a sign of success in the mission to express ``the supremacy of pure feeling or perception in the pictorial arts'' \cite{Danchev2011}?

Noticing the discrepancy in performance of the computational complexity measures for different image categories, \cite{Saraee2018} suggested that the metrics are not generalizable, and proposed creating specialized metrics for every domain. This echoes the suggestion that complexity may not be a uni-dimensional measure \cite{Oliva2004}. The approach we introduce here and its implementation in MSSC opens the venue for formulating testable hypotheses about special concepts of complexity for each case. 

The MSSC's struggles with man-made images, especially the categories of art and infographics, did not end at the lower spatial frequencies. The correlation between subjective and partial complexity at mid-scales was also numerically lower than for the other categories. It is plausible that this could be attributed to the lack of general cultural knowledge humans use in viewing these types of images. In semiotic terms, when arbitrary, symbolic signs are present, their purely visual complexity will not reflect the complexity of the derived interpretation. An example of this is shown in Figure 10. For such images, complexity arises in the space of interpretations, not visual composition.

The present study was limited in that we could not quantify the agreement between human raters in their complexity judgements. It is possible that subjective complexity for images that rely on cultural knowledge would be more variable between the raters, which would naturally diminish the correlations with objective complexity. Future studies may address this issue. In addition, it is possible that humans can focus on perceptual - as opposed to conceptual - complexity when instructed to do so \cite{Madan2018}. In this case, we would expect MSSC to yield higher correlations with subjectively rated visual complexity, while we would not expect any formalized measure to surpass subjective ratings of conceptual complexity.

\subsection{Future directions}

\subsubsection{Informational and effective complexity}

\begin{figure}[hbt!]
  \centering
  \includegraphics[width=12cm]{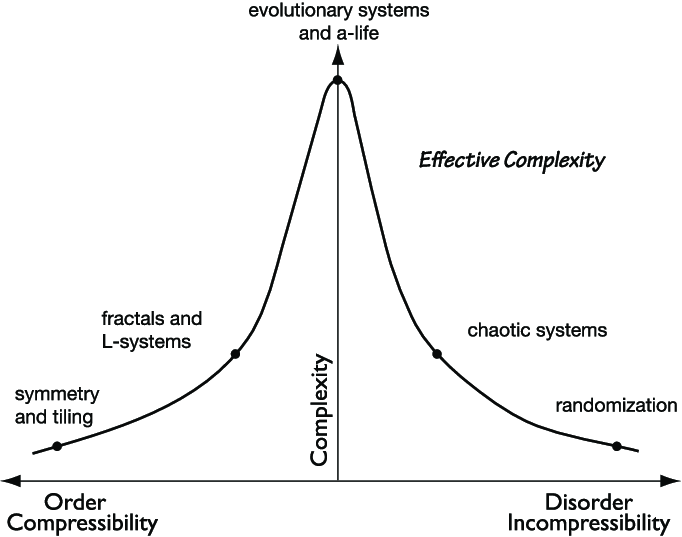}

  \caption{Informational and effective complexity. Image taken from Galanter, 2019.}
\label{complexity_types}
\end{figure}

As reviewed in the introduction, complexity affects aesthetic judgements and information exploration in a nonlinear manner \cite{Poli2024,Oudeyer2018,Berlyne1971}. It could be, however, that the non-linearity is driven by the fact that complexity is usually measured as the amount of information, while humans intuitively define it as "effective complexity" \cite{GellMann1996}, Fig.~\ref{complexity_types}. The former tends to be maximized at complete chaos/randomness, while the latter is maximized in the state of balance between order and randomness. With this distinction in mind, it seems sensible to revisit existing studies, testing whether the U-curve rule still holds when an effective complexity measure is applied instead of an informational one. Now that we have shown its correlation to human ranking, MSSC provides an opportunity for that.

\subsubsection{Physiological plausibility}

MSSC is the first complexity measure that can implement different methods of information processing at the lowest level, which would correspond to the level of the retina, LGN, and V1 in human vision. In the present study we used Fourier transform as a consensus approximation. Future studies could test other methods, such as wavelet analysis, and test the impact of different approaches to colour processing.  

\subsubsection{Practical applications}

MSSC holds practical utility concerning design and infographics, particularly in studies exploring visual strain versus informational value within design contexts.

It has been shown that humans have an aesthetic preference for natural scenes \cite{Kaplan1972} and that low-level visual features and spatial properties have a significant effect on aesthetic perception of scenes \cite{Kardan2015}. Considering the difference in the impact of spatial scales for man-made and natural images, studying the distribution of partial complexity could lead to clear guidelines or even automated testing of design and infographics.

\subsubsection{Conclusions}

To summarize, compared to alternative approaches, MSSC offers advantages in terms of consistency of complexity metric across categories of images, computational feasibility, and physiological plausibility. It holds the potential to inform future research in curiosity, attention, and aesthetic perception. Most importantly, it opens the venue for generating testable hypotheses about other, non-unitary conceptualizations of complexity.

\begin{appendices}

\section{Coarse graining procedure and multi-scale complexity}
\label{appendix-coarse-graining}

To illustrate the coarse graining procedure, let us focus on the case of grayscale images. Such an image can be viewed as an intensity function $f(x)$ defined on some domain $D$ (usually rectangular canvas), with $f=1$ corresponding to white color, and $f=-1$ -- to black color. The image can be coarse-grained by convolving $f(x)$ with some scale-dependent filter $g_{d\lambda} (x)$ to remove the features and details of size smaller than $d\lambda$: 

\begin{equation}
    f_{d\lambda}(x) = \int\limits_D f(x) g_{d\lambda}(x).
\end{equation}
If $f_{d\lambda}(x)$ is nearly identical to $f(x)$, it means that by removing small features from the pattern, we did not lose essential information, and there are no structures associated with scale $d\lambda$. In this case, we claim that this scale does not contribute to structural complexity. On the other hand, if the difference between $f(x)$ and $f_{d\lambda}(x)$ is substantial, it is a situation similar to observed in real complex systems, when two different scales have rather distinct properties, and contribution of scale $d\lambda$ to structural complexity can be regarded as large. To quantify this dissimilarity in the MSSC approach, non-normalized overlap is introduced:
\begin{equation}
\langle f(x) | f_{d\lambda}(x) \rangle \equiv \int\limits_D f(x) \cdot f_{d\lambda}(x). 
\end{equation}
 Only the original pattern is normalized, so that $\langle f(x)| f(x) \rangle=1$. Normalizing subsequent scales does not make much sense because the overall change in intensity over the course of coarse graining is also a part of overall dissimilarity between different scales, and must not be ignored.
 
 The convolution procedure is subsequently applied a number of times with the same smearing parameter $d\lambda$, and the partial complexity associated with scale $kd\lambda$ is defined as 
\begin{gather}
   {\cal C}_k= |\langle f_{kd\lambda}(x)|f_{(k+1)d\lambda}(x)\rangle - \frac12\left( \langle f_{kd\lambda}(x)|f_{kd\lambda}(x)\rangle + \langle f_{(k+1)d\lambda}(x)|f_{(k+1)d\lambda}(x)\rangle\right)|=\nonumber \\
   \frac{1}{2} \int_D \left( f_{(k+1)d\lambda}(x) -f_{k d\lambda}(x) \right)^2 dx,
   \nonumber
\end{gather} 
Summing up this over all scales, one obtain a number called {\it multi-scale structural complexity} $\cal C$:
\begin{equation}\label{eq:C_definition}
    {\cal C} = \sum\limits_k {\cal C}_k.
\end{equation}
Both $\cal C$ as an integral characteristic accounting for features emerging at every new scale, and the set of partial complexities $\{{\cal C}_k\}$ contain important information about the pattern, and are usually used together.

In the case of color patterns, partial and overall complexities in each RGB channel are computed independently and then averaged. Each channel is scaled to $[-1,1]$ range, with $-1$ representing the complete absence of the corresponding color, and $+1$ -- its maximal contribution.

%\section{Comparison of coarse graining approaches}
%\label{appendix-wavelets}

%In addition to Fourier transform, we tried two other methods of performing coarse graining: Gaussian blur and wavelet transform.

%\begin{figure}[hbt!]
 % \centering
  %\includegraphics[width=16cm]{cg_shapes.png}

  %\caption{Gaussian blur introduces distortions during later steps of coarse graining.}
%\label{cg_steps_shapes}
%\end{figure}

%\begin{figure}[hbt!]
 % \centering
  %\includegraphics[width=16cm]{cg_lineart.png}

  %\caption{Wavelets preserve shape at the expense of luminocity which is incompatible with the current pairwise comparison process.}
%\label{cg_steps_lineart}
%\end{figure}

\section{MSSC without spatial scale filtering}
\label{appendix-scale}

\begin{table}[!h]
\centering
\begin{tabular}{|l|c|c|}
\hline
 & MSSC & \begin{tabular}[c]{@{}c@{}}MSSC\\ (middle scale)\end{tabular} \\ \hline
scenes & 0.54 & 0.62 \\ \hline
objects & 0.34 & 0.46 \\ \hline
suprematism & 0.58 & 0.76 \\ \hline
interior design & 0.52 & 0.60 \\ \hline
advertisements & 0.44 & 0.52 \\ \hline
art & 0.05 & 0.36 \\ \hline
infographics & -0.02 & 0.38 \\ \hline
\end{tabular}
\vskip5pt
\caption{MSSC version incorporating only middle scales showed significantly better correlation with human ranking than MSSC computed across all spatial scales.}
\end{table}

%\section{impact of coarse graining detalisation}
%\label{appendix-cg-steps}

%\begin{figure}[hbt!]
 % \centering
  %\includegraphics[width=16]{}

  %\caption{more coarse graining steps doesn't improve the outcome}
%\label{cg_steps}
%\end{figure}

%\section{Outliers}
%\label{outliers}

%\begin{figure}[hbt!]
 % \centering
  %\includegraphics[width=16]{}

  %\caption{examples}
%\label{cg_steps}
%\end{figure}

\end{appendices}

\bibliography{jov_paper}
\bibliographystyle{unsrt}

\end{document}